\documentclass[prl,twocolumn,showpacs,amsmath,amssymb,superscriptaddress]{revtex4-1}

\pdfoutput=1
\usepackage[usenames,dvipsnames]{color}
\usepackage[colorlinks=true, linkcolor=BrickRed, citecolor=Blue, urlcolor=Blue, filecolor=Blue]{hyperref}
\usepackage{longtable}
\usepackage{epsfig}
\usepackage{amssymb}
\usepackage{amsmath}
\usepackage{graphicx}
\usepackage{verbatim}
\usepackage{xspace}

\def\ijmpa{Int.\ J.\ Mod.\ Phys.\ A}

\def\araa{ARA\&A}%
\def\apj{ApJ}%
\def\apjl{ApJ}%
\def\apss{Ap\&SS}%
\def\aap{A\&A}%
\def\jcap{JCAP}%
\def\mnras{MNRAS}%
\def\prd{Phys.\ Rev.\ D}%
\def\prl{Phys.\ Rev.\ Lett.}%
\def\nat{Nature}%
\def\science{Science}%
\def\physrep{Phys.\ Rep.}%
\def\plb{Phys. Lett. B}%
\definecolor{darkblue}{rgb}{0,0,0.5}
\definecolor{darkgreen}{rgb}{0.1,0,0.3}
\definecolor{darkred}{rgb}{0.6,0,0}

\newcommand{\nc}{\newcommand}

\nc{\ba}{\begin{eqnarray}}
\nc{\ea}{\end{eqnarray}}
\nc{\ga}{\gamma}
\nc{\om}{\omega}

\nc{\x}{{\bf x }}
\nc{\mx}{m_\chi}
\nc{\mnuc}{m_{N}}
\nc{\ud}{\,\mathrm{d}}
\nc{\ssi}{\sigma_{\mathrm{SI}}}
\nc{\ssd}{\sigma_{\mathrm{SD}}}
\nc{\tq}{\tilde \q}
\nc{\dmin}{\delta_{\mathrm{min}}}
\nc{\dmax}{\delta_{\mathrm{max}}}
\nc{\DS}{\textsf{DarkStec}\xspace}
\nc{\GS}{\textsf{GARSTEC}\xspace}
\nc{\dstars}{\textsf{DarkStars}\xspace}
\nc{\vrel}{v_{\rm rel}}
\nc{\qref}{q_{\rm ref}}
\nc{\sv}{\langle \sigma v \rangle_{\rm ann}}
\nc{\er}{E_\mathrm{R}}
\begin{document}

\title{A possible indication of momentum-dependent asymmetric dark matter in the Sun}

\author{Aaron C. Vincent}
 \affiliation{Institute for Particle Physics Phenomenology (IPPP), Department of Physics, Durham University, Durham DH1 3LE, UK. \href{mailto:aaron.vincent@durham.ac.uk}{\normalfont aaron.vincent@durham.ac.uk}}

\author{Pat Scott} 
 \affiliation{Department of Physics, Imperial College London, Blackett Laboratory, Prince Consort Road, London SW7 2AZ, UK. \href{mailto:p.scott@imperial.ac.uk}{\normalfont p.scott@imperial.ac.uk}} 

\author{Aldo Serenelli}
 \affiliation{Institut de Ci\`encies de l'Espai (CSIC-IEEC), Campus UAB, Carrer de Can Magrans, s/n
08193 Cerdanyola del Vall\'es, Spain. \href{mailto:aldos@ice.csic.es}{\normalfont aldos@ice.csic.es}}

\begin{abstract}
{Broad disagreement persists between helioseismological observables and predictions of solar models computed with the latest surface abundances.  Here we show that most of these problems can be solved by the presence of asymmetric dark matter coupling to nucleons as the square of the momentum $q$ exchanged in the collision. We compute neutrino fluxes, small frequency separations, surface helium abundances, sound speed profiles and convective zone depths for a number of models, showing more than a $6\sigma$ preference for $q^2$ models over others, and over the Standard Solar Model.  The preferred mass (3\,GeV) and reference dark matter-nucleon cross-section ($10^{-37}$\,cm$^2$ at $q_0 = 40$\,MeV) are within the region of parameter space allowed by both direct detection and collider searches.}
\end{abstract}

\maketitle

\textbf{\textit{Introduction.---}}
Since the downwards revision of the solar photospheric metallicity \cite{APForbidO,*CtoO,*AspIV,*AGSS,*AGSS_NaCa,*AGSS_FePeak,*AGSS_heavy}, a number of discrepancies have appeared between models of the solar interior and helioseismology.  Models computed with the revised photospheric abundances show poor agreement with the observed depth of the convection zone, sound speed profile, surface helium abundance and small frequency separations \cite{Basu:2004zg, *Bahcall:2004yr, *Basu08, Serenelli:2009yc}.  A number of explanations have been proposed \cite{Guzik05,*Drake05,*Charbonnel05,*Christensen09,Serenelli11}, some based on axion-like particles \cite{Vincent12} or modified energy transport in the solar interior due to dark matter (DM) \cite{Taoso10,Frandsen10,*Cumberbatch10}, but none has proven compelling.

Here we demonstrate that the existence of weakly-interacting asymmetric dark matter (ADM; \cite{Petraki13,*Zurek14}) with a mass of a few GeV can explain most of these anomalies, if (and only if) the strength of the interaction between DM and nucleons depends on the momentum $q$ exchanged between them.  In particular, we find a more than $6\sigma$ preference for a coupling proportional to $q^2$.  Unlike weakly-interacting massive particles (WIMPs), the motivation for ADM comes from the baryonic sector of the standard model, and relies on an initial asymmetry between DM and anti-DM to generate the correct relic abundance. Crucially, this can lead to an absence of self-annihilation today, allowing large quantities of ADM to accumulate in stars like the Sun.

\textbf{\textit{Momentum-dependent dark matter.---}}
The scattering cross-section between DM and nucleons can depend on both the relative velocity of the colliding particles ($\vrel$) and the momentum that they exchange ($q$).  The first term in series expansions of the cross-section is independent of both $\vrel$ and $q$, and dominates in models such as supersymmetry.  In other models, this term is suppressed, and the leading contribution comes from terms with a non-trivial dependence on $\vrel$ or $q$ \cite{Pospelov00,*Sigurdson04}.  At low masses, such a dependence has been one of the theoretical mechanisms proposed to reconcile various anomalies in direct searches for dark matter \cite{Chang:2009yt,*Feldstein:2009tr,*Feldstein10,*Fitzpatrick10,*Frandsen:2013cna,*DelNobile:2014eta,*Cherry:2014wia,*Arina:2014yna}.

Here we focus on an effective spin-independent (SI) elastic cross-section between DM $\chi$ and nucleons of the form
\begin{equation}
\sigma_{\chi-{\rm nuc}} = \sigma_0 \left(\frac{q}{q_0}\right)^2,
\end{equation}
where $q_0$ is a reference momentum used to normalise the scattering cross-section; we choose $q_0 = 40$\,MeV, which corresponds to a typical nuclear recoil energy of $\sim$10\,keV in direct detection experiments.  Such a $q^2$ SI form to the cross-section can arise from, e.g. effective DM-quark operators like $\bar\chi\gamma_5\chi\bar qq$ and $\bar\chi\sigma_{\mu\nu}\gamma_5\chi\bar q\sigma^{\mu\nu}q$ \cite{Goodman10,*Kumar13}. The former operator is particularly appealing in its simplicity, arising from the exchange of a pseudoscalar mediator.

\textbf{\textit{Helioseismology and dark matter.---}}
The impacts of DM-nucleon scattering on helioseismology and stellar structure have been well studied \cite{Taoso10,Frandsen10,*Cumberbatch10,Lopes14,*Lopes:2014,Lopes02a,*Bottino02,*Lopes:2012,*Iocco12}.  Weakly-interacting DM from the Galactic halo is captured when it passes through the Sun, scatters onto a bound orbit \cite{Gould87b}, undergoes repeated additional scattering and energy loss, and eventually settles into the solar core.  DM-nucleon scattering provides an additional means of conductive energy transport: DM particles absorb energy in the hottest, central part of the core, then travel to a cooler, more distal region before scattering again and redepositing their energy \cite{GouldRaffelt90a}.  This decreases the temperature contrast over the core region and reduces the central temperature. The cooler core produces fewer neutrinos from the most temperature-sensitive fusion reactions, so the $^8$B and $^7$Be neutrino fluxes observed at Earth can be noticeably reduced. This is accompanied by a smaller increase in the $pp$ and $pep$ fluxes, as required by the constancy of the solar luminosity.

The structural changes in the core shift the balance between gravity and pressure elsewhere, leading to global readjustments in models constrained to fit the solar radius $R_\odot$, luminosity $L_\odot$ and metal to hydrogen abundance ratio $(Z/X)_\odot$ at the solar system age $t_\odot$.  A widely used seismic diagnostic, the depth of the solar convective envelope $R_{CZ}$, is determined by the temperature gradient immediately below the convective envelope. In our DM models, the gradient in this region is slightly steeper than in the Standard Solar Model (SSM), leading to a modest but measurable deepening of the convective envelope.  The lower core temperature leads to lower nuclear fusion rates, which must be compensated for by increasing the hydrogen abundance so that the integrated nuclear energy release accounts for $L_\odot$. The initial helium mass fraction and the present day surface value $Y_s$ are thus lower in models where DM contributes to energy transport.  In general, helioseismic diagnostics are affected by changes in temperature ($T$), mean molecular weight ($\bar\mu$), and their gradients, as the solar sound speed varies as $\delta c_s/c_s \approx \frac{1}{2} \delta T/T - \frac{1}{2} \delta \bar\mu/\bar\mu$ (neglecting here a small term from variation of the adiabatic index $\Gamma_1$). If $\nu_{n,\ell}$ is the frequency corresponding to the eigenmode of radial order $n$ and angular degree $\ell$, then the so-called frequency ratios
\begin{align}
r_{0,2} = \frac{\nu_{n,0} - \nu_{n-1,2}}{\nu_{n,1}-\nu_{n-1,1}} \ \ {\rm and} \ \ 
r_{1,3} = \frac{\nu_{n,1} - \nu_{n-1,3}}{\nu_{n+1,0}-\nu_{n,0}}, 
\end{align}
are given by
\begin{align}
r_{\ell,\ell+2}(n) \approx -(4\ell+6) \frac{1}{4\pi^2 \nu_{n,\ell}}\int_0^{R_\odot}{\frac{dc_s}{dR}\frac{dR}{R}},
\end{align}
for $n \gg 1$. These are weighted towards the core, so give information on the central region of the Sun \cite{Basu:2006vh}. In this work we use solar data from BiSON \cite{bison2007}, from which ratios can be computed for $n>8$.

The major technical advance here over earlier work \cite{Taoso10,Frandsen10,*Cumberbatch10,Lopes14,*Lopes:2014} is that we compute solar models using an accurate treatment of energy transport and solar capture by momentum-dependent DM-nucleon interactions.  The correct transport treatment is quite involved \cite{VincentScott2013}.  The capture rate of $q^2$-dependent DM by the Sun is \cite{Vincent14} 
\begin{align}
&C_\odot(t) = 4\pi \int_0^{R_\odot} R^2 \int_0^\infty \frac{f_\odot(u)}{u} w^2 \sum_i \sigma_{N,i} n_i(R,t) \frac{\mu_{i,+}^2}{\mu_i}\nonumber\\
&\hspace{4mm}\times \Theta\left(\frac{\mu_i v_{\rm esc}^2(R,t)}{\mu_{i,-}^2} - u^2 \right) \left(\frac{m_\chi w}{q_0}\right)^{2} I_{\rm FF} \ud u\ud R,
\label{caprate}
\end{align}
where $R_\odot$ is the solar radius, $m_\chi$ the DM mass, $v_{\rm esc}(R,t)$ the local escape speed at height $R$ in the Sun, $f_\odot(u)$ the distribution of halo DM particle speeds $u$ in the solar frame, $w \equiv \sqrt{u^2 + v_{esc}^2}$, $\sigma_{N,i}$ and $n_i$ are the DM-nucleus scattering cross-section and local number density respectively for nuclear species $i$, $\mu_i \equiv \mx/m_{N,i}$, $\mu_{i,\pm} \equiv (\mu_i \pm 1)/2$, and $I_{\rm FF}$ is the form factor integral.  For hydrogen,
\begin{equation}
I_{\rm FF} = \frac{\mu_{{\rm H},+}^2}{2\mu_{\rm H}^2} \left[\frac{\mu_{\rm H}^2}{\mu_{{\rm H},+}^4} - \frac{u^4}{w^4}\right].
\label{ffHresult}
\end{equation}
For heavier elements, assuming a Helm form factor gives
\begin{equation}
I_{\rm FF} = \frac{\mu_i}{(B_i \mu_i)^2}\left[ \Gamma\left(2,B_i\frac{u^2}{w^2}\right) - \Gamma\left(2,B_i\frac{\mu_i}{\mu_{i,+}^2}\right) \right],
\label{ffheavyresult}
\end{equation}
with $\Gamma(m,x)$ the upper incomplete gamma function. Here $B_i\equiv \frac12 \mx w^2 / E_i$, where $E_i$ is a constant given in \cite{Gould87b} for each nuclear species.

\textbf{\textit{Simulations of $q^2$\,ADM in the Sun.---}}
To study the impact of $q^2$ ADM on solar observables, we merged the solar structure and dark stellar evolution codes \GS \cite{weiss:2008, Serenelli11} and \dstars \cite{Scott09,*Scott09b}, then implemented momentum-dependent transfer as per \cite{VincentScott2013} and capture as in Eq.\ (\ref{caprate}), creating a precision dark solar evolution package \DS.    We computed solar models matching $(Z/X)_\odot$, $R_\odot$ and $L_\odot$ at the solar age $t_\odot$ over a grid of ADM masses and cross-sections $\sigma_0$, for regular SI and SD (spin-dependent) ADM, as well as $q^2$ momentum-dependent SI ADM.  We assumed passage of the Sun at 220\,km\,s$^{-1}$ through a standard Maxwell-Boltzmann halo with velocity dispersion 270\,km\,s$^{-1}$ and local DM density 0.38\,GeV\,cm$^{-3}$.  On the basis of the observed $^8$B and $^7$Be neutrino fluxes, depth of the convection zone, surface helium fraction and sound speed profile, we selected the best-fit model within each of these grids: for \{SD, SI, $q^2$ SI\} models, $m_\chi = \{5,5,3\}$\,GeV and $\sigma_0 = \{10^{-36},10^{-34},10^{-37}\}$\,cm$^2$.

\begin{figure}[tb]
\includegraphics[width=.5\textwidth]{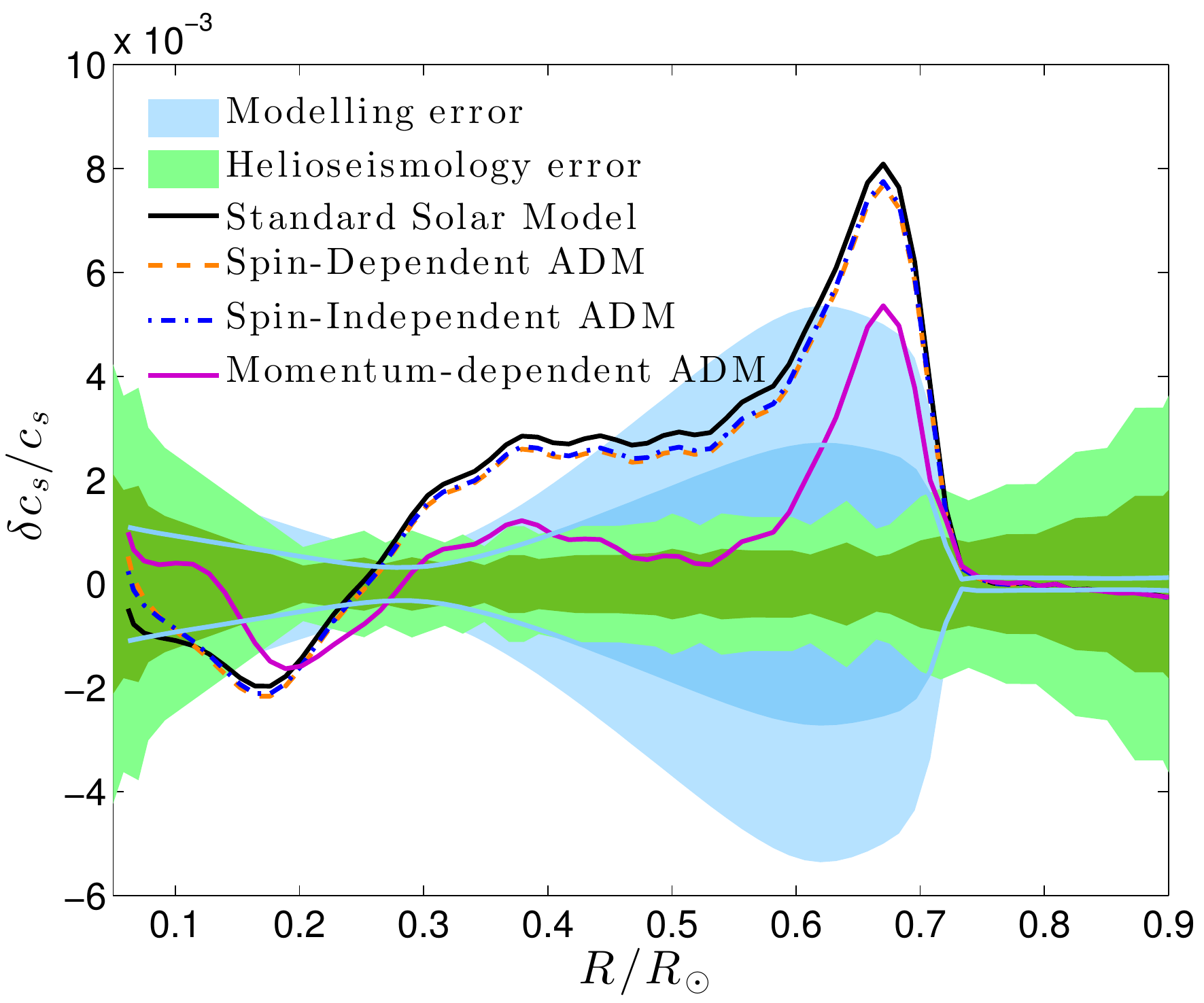}
\caption{Deviation of the radial sound speed profile ($\mathrm{Sun}-\mathrm{model}$)/Sun in the solar interior from the values inferred from helioseismological data, for the Standard Solar Model (SSM) and three models of asymmetric dark matter (ADM). Coloured regions indicate 1 and 2$\sigma$ errors in modelling (thick blue band) and on helioseismological inversions \cite{1997APh.....7...77D,*2001PhLB..503..121F} (thinner green band). The combination ($m_\chi,\sigma_{\chi-nuc}$) for each model is chosen to give the best overall improvement with respect to the SSM.}
\label{fig1}
\end{figure}

\begin{figure*}[tb]
\begin{minipage}{.465\textwidth}
\includegraphics[width=\linewidth]{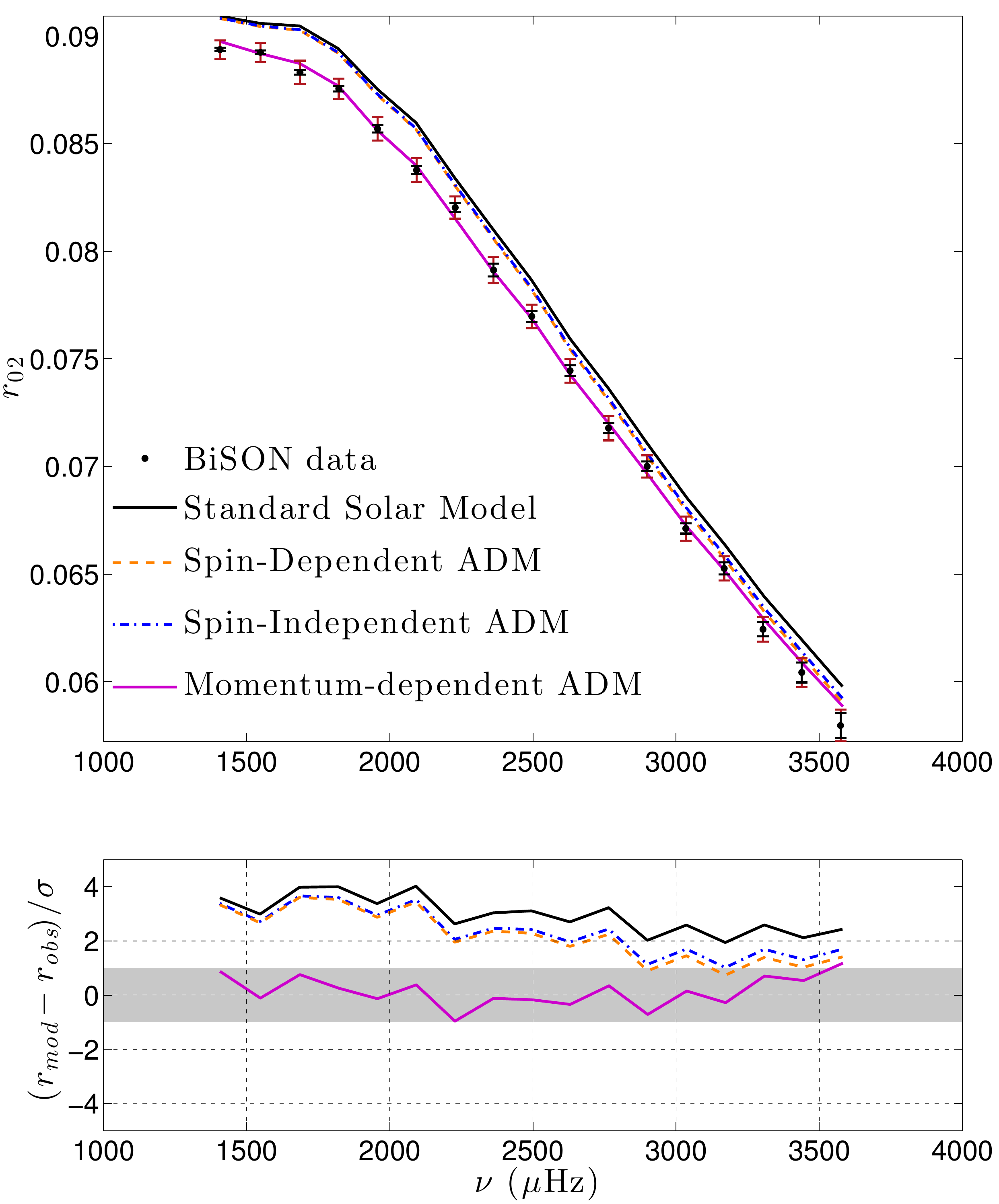}
\end{minipage}
\begin{minipage}{.05\textwidth}
\hspace{\linewidth}
\end{minipage}
\begin{minipage}{.465\textwidth}
\includegraphics[width=\linewidth]{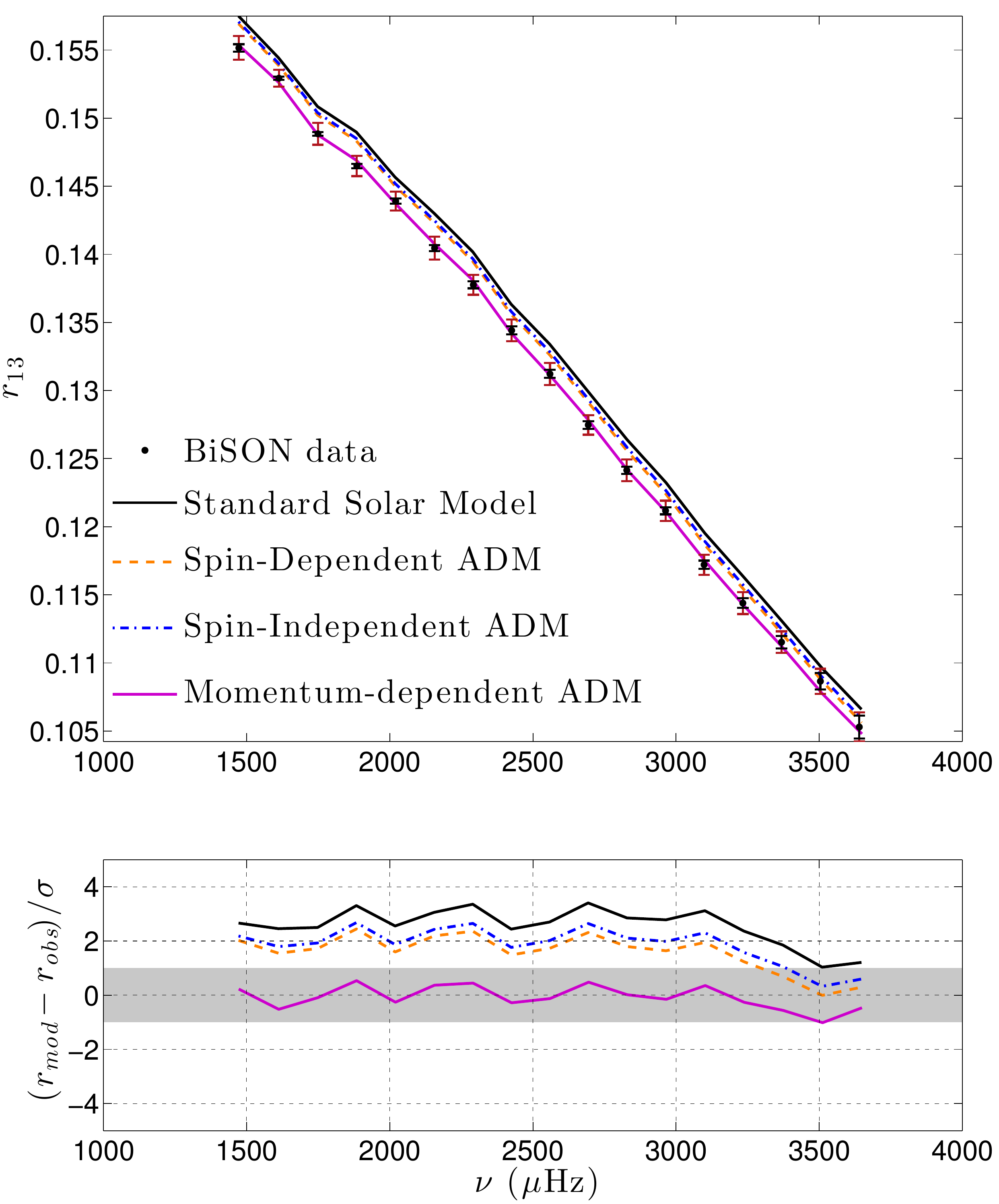} 
\end{minipage}
\caption{Small frequency separations $r_{02}$ (\textit{left}) and $r_{13}$ (\textit{right}), for the SSM, regular SI and SD ADM, and $q^2$ SI momentum-dependent ADM. Data from BiSON \cite{bison2007} show observational (inner black) and total errors including modelling uncertainties (outer red).  Bottom panels give residuals as number of standard deviations from the observed values; grey bands are $\pm$$1\sigma$.}
\label{fig2}
\end{figure*}

In Fig.\ \ref{fig1} we compare the sound speed profile predicted by each of the three best-fit models to that inferred from helioseismic inversions (as presented in \cite{Serenelli:2009yc}). We also show the profile of the SSM as computed with the most up to date input physics described in \cite{Serenelli11} and with the latest photospheric abundances \cite{AGSS}. This model is an update to the AGSS09ph model of \cite{Serenelli:2009yc}. Modeling errors are computed by propagating the errors of the input parameters to each observable by using power-law expansions \cite{Serenelli2013, *Villante:2013mba,*vinyoles:2015}.

SI and SD ADM provide little improvement over the SSM.  Momentum-dependent ADM significantly improves agreement with the observed sound speed profile, both at the base of the convection zone and in the outer part of the core, bringing the discrepancy down to little more than $\sim$$2\sigma$.  Momentum-dependent ADM evacuates energy from the solar core, causing it to become cooler, in turn increasing the central hydrogen fraction and reducing the mean molecular weight of the core material.  The net effect is a decrease in the sound speed. At intermediate regions, where DM deposits energy, the temperature is slightly higher, forcing a steeper temperature gradient towards the bottom of the convective envelope, and therefore a deeper $R_{\rm CZ}$.

We also computed the small frequency separations $r_{02}$ and $r_{13}$ (Fig.\ \ref{fig2}).  The agreement of predictions from momentum-dependent DM with the observed ratios is remarkable, barely passing beyond a single standard deviation for any ratio.  None of the other models is able to produce a remotely competitive fit.  

In Table \ref{tab} we give the neutrino fluxes, $R_{\rm CZ}$ and $Y_{\rm s}$ predicted by all models, along with contributions to a global $\chi^2$ statistic from each.  In all models, $pp$ neutrino fluxes are affected by less than $0.1\sigma$ \cite{Bellini:2014uqa}, so we do not include them.  Assuming Gaussian errors, the $q^2$ model yields a p-value of $0.85$, indicating an excellent overall fit to data.  All the other models have $p<10^{-10}$, indicating that they are ruled out with greater than $6\sigma$ confidence. 

We see that although the $q^2$ model gives slightly worse agreement with the observed neutrino fluxes and $Y_{\rm s}$ than the SSM, the overall fit is dramatically better. The fit to $R_{\rm CZ}$ is improved from a $2.2\sigma$ discrepancy in the SSM to little more than a standard deviation.  The largest contributor to the global $\chi^2$ of the $q^2$ model is $Y_{\rm s}$, which changes from the SSM as $0.2356\rightarrow0.2327$ (a $2.6\sigma\rightarrow3.2\sigma$ discrepancy).  

We include $r_{\ell,\ell+2}$ but not $c_s$ in the $\chi^2$, as the former is more precise, and the two datasets are strongly correlated. Different $r_{\ell,\ell+2}$ values are also correlated. For the data that we use, however, the correlation is $<1\%$ between different $n$ and $<8\%$ between $r_{02}$ and $r_{13}$.  We hence include all points in the $\chi^2$.  Using e.g. $r_{02}$ only (which gives a worse fit than $r_{13}$) would only reduce $p$ to 0.18 -- still an excellent fit.  

The $q^2$ model also yields a `parameter goodness of fit' \cite{MaltoniSchwetz} of 0.30, indicating that the degree of tension between different observables in this model is quite acceptable, at barely more than $1\sigma$.  For this calculation, we have $\bar\chi^2 = 11.8$ with 10 degrees of freedom, conservatively treating each of $r_{02}$ and $r_{13}$ as a single independent observable; were we to instead treat each frequency as a degree of freedom, the corresponding $p$ value would be even better.

The principal difference between the ADM models we consider here is the effect on the DM mean free path $\ell_\chi$. In all cases, $\ell_\chi$ rises rapidly away from the dense solar core. This rise occurs more rapidly with $r$ for SI than SD scattering. This larger gradient allows SI ADM to transport energy much farther away from the core than SD ADM. When the coupling is proportional to $q^2$, there is a further enhancement of the mean free path that goes as $(q_0/m v_T)^2$, where $v_T^2 \sim T$ is the typical nuclear thermal velocity. This facilitates even more energy deposition at higher radii, yielding the vast improvements in $r_{CZ}$, $r_{02}$, $r_{13}$ and $c_s(r)$ that we see. Although $q^2$ couplings suppress the capture rate, this is not enough to suppress the effects of conduction, as in the case of a $q^4$ coupling \cite{Vincent14}. The full details of the thermal conduction calculation are given in \cite{VincentScott2013}.

\textbf{\textit{Discussion.---}}
This is the first real exploration of the effects of momentum-dependent dark matter on solar physics.  Previous papers dealt with regular SI and SD couplings \cite{Taoso10,Frandsen10,*Cumberbatch10,Lopes02a,*Bottino02,*Lopes:2012,*Iocco12}, but of those only \cite{Taoso10} included the correct treatment of conductive energy transport by DM.  Accounting for (small) improvements in the underlying solar modelling here relative to \cite{Taoso10}, our SI and SD results are in good agreement with their findings.  The only other investigations to date of non-standard couplings in the context of helioseismology \cite{Lopes14,*Lopes:2014} involved approximate treatments of mixed $q$-$\vrel$-dependent cross-sections as purely $\vrel$, without proper capture or transport calculations, nor consideration of all observational consequences.  A $\vrel^{-2}$ SD cross-section, for example, can indeed provide improvements over the SSM in terms of $c_s$ and $R_{\rm CZ}$, but these are outweighed by more severe decreases in the $^8$B and $^7$Be neutrino fluxes \cite{Vincent14}.

The mass (3\,GeV) and cross-section ($10^{-37}$\,cm$^2$) of $q^2$ momentum-dependent DM preferred by solar physics are in agreement with bounds from direct searches \cite{Guo2013}, and are even tantalisingly close to some of the preferred regions in analyses of direct detection anomalies \cite{Chang:2009yt,*Feldstein:2009tr,*Feldstein10,*Fitzpatrick10,*Frandsen:2013cna,*DelNobile:2014eta,*Cherry:2014wia,*Arina:2014yna}.  Models with appropriate couplings (e.g. $\bar\chi\gamma_5\chi\bar qq$) are also still allowed by collider searches \cite{Cheung12}, so the prospects for soon confirming or refuting the existence of $q^2$ ADM resembling our best-fit model appear favourable. 

\begin{table}[tb]
\caption{Measured and predicted solar observables.}
\label{tab}
\begin{tabular}{ l r  r  r  r  r  r  r }
\hline\hline \vspace{-4mm}\\
                                                                  &  SSM  &  SD      &  SI      &  $q^2$ SI & Obs.\footnote{Neutrino data and obs. errors inferred from Borexino data \cite{Serenelli11}.} & $\sigma_{\rm obs}$  & $\sigma_{\rm model}$ \\      
\hline
$\phi_{\nu}^{^8\mathrm{B}}$ \footnote{In units of $10^6$ cm$^{-2}$s$^{-1}$.}
                                                                  & 4.95  &  4.39    &  4.58    &  3.78     & 5.00   & 3\%                 & 14\%                 \\
$\phi_{\nu}^{^7\mathrm{Be}}$ \footnote{In units of $10^9$ cm$^{-2}$s$^{-1}$.}
                                                                  & 4.71  &  4.58    &  4.62    &  4.29     & 4.82   & 5\%                 & 7\%                  \\
$ R_{\rm CZ}/R_\odot$                                 & 0.722 &  0.721   &  0.721   &  0.718    & 0.713  & 0.001               & 0.004                \\
$Y_{\rm s}$                                                 & 0.2356&  0.2351  &  0.2353  &  0.2327   & 0.2485 & 0.0034              & 0.0035               \\
$\chi^2_{^8\mathrm{B}}$                               & 0.0   &  0.9     &  0.9    &  4.9      &        &                     &                      \\
$\chi^2_{^7\mathrm{Be}}$                              & 0.1   &  0.4     &  0.4     &  1.9      &        &                     &                      \\
$\chi^2_{R_{\rm CZ}}$                                   & 4.8   &  3.8     &  3.8     &  1.5      &        &                     &                      \\
$\chi^2_{Y_{\rm s}}$                                      & 7.0   &  7.5     &  7.3     &  10.5     &        &                     &                      \\
$\chi^2_{r_{02}}$                                           & 156.6 &   95.3   &  105.2   &   5.6     &        &                     &                      \\
$\chi^2_{r_{13}}$                                           & 119.3 &   50.7   &  67.2   &   3.1     &        &                     &                      \\
$\chi^2_{\rm total}$                                       & 287.8 &  158.5   &  185.2   &  27.5     &        &                     &                      \\
$p$                                                              & $<$$10^{-10}$ & $<$$10^{-10}$ & $<$$10^{-10}$ & 0.845          &        &                     &                      \\
\hline
\end{tabular}
\end{table}

\textbf{\textit{Acknowledgements.---}}
We acknowledge funding support from NSERC, FQRNT, STFC and European contracts FP7-PEOPLE-2011-ITN and PITN-GA-2011-289442-INVISIBLES, EPS2013-41268-R (MINECO) and 2014SGR-1458. Calculations were performed on SOM2 at IFIC funded by PROMETEO/2009/116 and FPA2011-29678.

\bibliography{CandO,CObiblio,AbuGen,solarDM}

\end{document}